\documentclass[runningheads]{svmult}

\usepackage{graphicx}  

\def\erg{~\rm{erg}}
\def\ergs{{\rm erg}~{\rm s}^{-1} }
\def\ergscm2{~{\rm erg}~{\rm s}^{-1}~{\rm cm}^{-2} }
\def\gcm2{~{\rm g}~{\rm cm}^{2} }
\def\GeV{~\rm{GeV}}
\def\MeV{~\rm{MeV}}
\def\keV{~\rm{keV}}
\def\eV{~\rm{eV}}
\def\G{\rm{G}}
\def\K{\rm{K}}
\def\s{~\rm{s}}
\def\cm{\rm{cm}}

\def\lambdabar{\mathrel{\lower 1pt\hbox{$\mathchar'26$}\mkern-9mu
        \hbox{$\lambda$}}}

\def \eps {{\cal \scriptstyle E}}
\def \epr {\epsilon^\prime}
\def \eb  {\epsilon_{\scriptscriptstyle B}}

\def \st  {\sigma_{\scriptscriptstyle T}}
\def \skn  {\sigma_{\scriptscriptstyle KN}}

\def \sss {\scriptscriptstyle}

\newbox\grsign \setbox\grsign=\hbox{$>$} \newdimen\grdimen \grdimen=\ht\grsign
\newbox\simlessbox \newbox\simgreatbox \newbox\simpropbox
\setbox\simgreatbox=\hbox{\raise.5ex\hbox{$>$}\llap
     {\lower.5ex\hbox{$\sim$}}}\ht1=\grdimen\dp1=0pt
\setbox\simlessbox=\hbox{\raise.5ex\hbox{$<$}\llap
     {\lower.5ex\hbox{$\sim$}}}\ht2=\grdimen\dp2=0pt
\setbox\simpropbox=\hbox{\raise.5ex\hbox{$\propto$}\llap
     {\lower.5ex\hbox{$\sim$}}}\ht2=\grdimen\dp2=0pt
\def\simgreat{\mathrel{\copy\simgreatbox}}
\def\simless{\mathrel{\copy\simlessbox}}

\def\la{\mathrel{\hbox{\rlap{\hbox{\lower4pt\hbox{$\sim$}}}\hbox{$<$}}}}
\def\ga{\mathrel{\hbox{\rlap{\hbox{\lower4pt\hbox{$\sim$}}}\hbox{$>$}}}}

\begin{document}
\title*{Neutron Stars as Sources of High Energy Particles -- the case of RPP
\footnote{Invited Review at International School ``Physics and Astrophysics of Ultra High Energy
Cosmic Rays", June 2000, Paris-Meudon
(France), to be published in Lecture Notes in Physics (Springer)}}
\toctitle{Neutron Stars as Sources of High Energy Particles}
\titlerunning{Neutron Stars}
\author{Bronis{\l}aw Rudak}
\authorrunning{Bronis{\l}aw Rudak}

\institute{CAMK, Rabia{\'n}ska 8, Toru{\'n}, Poland\hfill\break
also TCfA NCU, Toru{\'n}, Poland}

\maketitle              

\begin{abstract}
Highly magnetised rapidly spinning neutron stars are widely considered 
to be natural sites for acceleration of charged particles. 
Powerful acceleration mechanism due to unipolar induction is thought to operate 
in the magnetospheres of isolated neutron stars, bringing the particles 
to ultrarelativistic energies at the expense of the neutron star rotational energy,
with inevitable emission of high energy photons.\hfill\break
The aim of this review is to present basic ingredients of modern models
of magnetospheric activity of rotation powered pulsars in the context
of high-energy radiation from these objects. 
Several aspects of pulsar activity are addressed and related to
spectacular results of pulsar observations with two major satellite
missions of the past -- CGRO and ROSAT.
It is then argued that high sensitivity experiments of 
the future - GLAST, VERITAS and MAGIC - will be vital for
a progress in our understanding of pulsar magnetospheric processes.\hfill\break
In a conservative approach rotation powered pulsars are not expected to be the sources 
of UHE Cosmic Rays. However, several scenarios have been proposed recently to 
explain the UHECR events above the GZK limit with the help of acceleration processes in the immediate 
surrounding of newly born pulsars. Major features of these scenarios are reviewed
along with references to contemporary models of magnetospheric activity.
\end{abstract}

\section{Introduction}
High energy radiation from various classes of galactic and extragalactic objects has been
observed for nearly 30 years. Large fraction of the galactic sources are 
associated with neutron stars: rotation powered pulsars (RPP), accretion powered pulsars (APP),
cooling neutron stars, and soft gamma-ray repeaters (SGR).
Rotation powered pulsars like Crab, Vela and Geminga have a long history of successful observations with baloon-born
and satellite gamma-ray and X-ray experiments. Performance of
old experiments had been, however, surpassed in terms of sensitivity, energy range, number of positive detections,
or photon statistics per object by the COMPTON Gamma-Ray Observatory (CGRO) and R{\" o}ntgen Satellit (ROSAT).  
Spectacular results of observational campaigns of RPP with ROSAT and CGRO induced a new wave of interest in
theoretical aspects of pulsar magnetospheric activity. 

Pair creation paradigm is a pivotal element in any model of magnetospheric activity of RPP.
Electron-positron pairs ($e^\pm$-pairs) are necessary since they
are thought to be responsible for radio emission observed in radiopulsars
and interpreted as the coherent curvature radiation of $e^\pm$ plasma.
Pairs can be produced in magnetospheric environments  either via photon absorption in a dense field
of soft photons (photon-photon collision) or via photon absorption in a strong magnetic field. 
In either case a supply of high-energy (HE) photons is required in order to fulfill stringent threshold 
conditions for the pair creation.
It it quite reasonable then to assume that not all of those HE-photons would be subject
to absorption. On the contrary, many HE photons will escape the magnetosphere without any attenuation.
This argument leads us to expect that RPP (and all radiopulsars in particular) should be
the sources of HE radiation.
To make the production of HE photons possible, highly relativistic charged particles are to be injected 
into the magnetosphere. One may speculate 
that some of these particles will either retain their energy
or regain it (under circumstances to be specified) upon escaping from the source.
It is up to theoretical models of the RPP activity to show whether a rate of HE radiation and/or particles
is interestingly high with respect to the sensitivity of recent and future HE detectors and telescopes.

This review focuses on RPP as sources of HE photons, presenting the most important
observational results as well as their interpretation in terms of basic processes expected in the magnetospheres of RPP.
The interpretation is offered by referring to a particular class of models of magnetospheric activity, 
known as polar gap models (or polar cap models). 
The name reflects the association of accelerator (the gap) with a polar cap on neutron star (NS) surface. 
Contrary to polar gap models, outer gap models \cite{ho} postulate the existence
of accelerators located in regions where local corotation
charge density reaches zero, close to the light cylinder. The $e^\pm$--pair creation
occurs there either via one photon magnetic absorption (Crab-type outer gaps)
or via photon-photon collisions (Vela type outer gaps).
These models are relevant for  both classical and millisecond pulsars with 
sufficiently high spin-down luminosity $L_{\rm sd}$.
The outer-gap accelerators cease to produce $e^\pm$ pairs once the pulsar crosses
the death-line $\log \dot P = 3.8\, \log P - 11.2$ \cite{chen}.
A modern version of the outer-gap accelerator, the so called `thick gap solution' \cite{cheng}, is however
able to accommodate pulsars of longer spin periods, like Geminga and B1055-52, removing at the same time
serious problems with the original model of \cite{ho}.
A detailed review of outer gap models in the context of HE radiation is available \cite{cheng}
and therefore we'll concentrate on polar gap models.

The review is organised in the following way: Sect.2 defines basic 
quantities and introduces the assumptions used in pulsar physics. 
The status of X-ray and gamma-ray observations of RPP and essential features of the radiation
detected from several sources are presented in Sect.3.
Sect.4 offers simple estimates of how effective a unipolar inductor (i.e. accelerator) can be when
acting in the framework of a neutron star. Sect.5 discusses vertical structure  
of some polar gap accelerators.
Sect.6 presents the properties of most important
radiative processes induced by such inductors inside the RPP magnetosphere. 
With energetic arguments formulated in Sect.4, Sect.7 addresses a question 
of whether  newly born and fastly spinning RPP might lead to generation of ultrarelativistic charged particles
responsible then for the UHECR events observed above the GZK limit. 
Sect.8 emphasizes the anticipated role of high sensitivity HE/VHE missions of 
the near future in contributing to the physics of RPP.

\section{Basic parameters}
The aim of this section is to define basic quantities used throughout the review
and to introduce their mutual relations.
Several excellent monographs covering this subject
in a detailed and sophisticated way are available e.g. \cite{michel} with a critical
discussion.

The starting point are  two major quantities measured 
for pulsars -- period $P$, interpreted as a period of rotation of a neutron star, 
and $\dot P$, its time derivative. Suppose, that a neutron star of radius $R_{\rm s}$ and
moment of inertia $I$ rotates with angular velocity $\Omega = 2\pi/P$ which
decreases in time (for whatever reason) at a rate $\dot \Omega = - 2\pi\, P^{-2}\dot P < 0$.
The rotational energy and its time derivative then read
\begin{eqnarray}\label{b1}
E_{\rm rot} &=& {1 \over 2}\, I\, {\Omega}^2 \simeq 2\times 10^{46}\, I_{45} P^{-2} {\erg} \\
\dot E_{\rm rot} &=& I\, {\Omega}\, \dot \Omega \simeq - 4\times 10^{31}\, I_{45} \dot P_{-15} P^{-3} {\ergs}
\end{eqnarray}
where $P$ is in seconds, $\dot P_{-15} \equiv \dot P / 10^{-15}$ and $I_{45} \equiv I/10^{45}\gcm2$. 
Instead of $\dot E_{\rm rot}$ one uses the so called spin-down luminosity $L_{\rm sd}$ defined as
\begin{equation}\label{b2}
L_{\rm sd} \equiv - \dot E_{\rm rot}.
\end{equation}
The name {\it luminosity} is misleading since the carriers of the major part of $\dot E_{\rm rot}$
are not luminous for us: no one has ever managed to ``see'' them 
in a direct way by any type of detector (but see Sect.7) and their nature is model-dependent, at least for the time being. 
Therefore we need a model for the spin-down of a neutron star.
Let us assume that a magnetic dipole is attached to the center of a neutron star, with
its moment inclined at angle $\alpha$ to the spin axis $\vec \Omega$, and let the mean
strength of the field at the stellar surface is $B_{\rm s}$. The magnetic dipole, rotating in 
a vacuum will emit energy at the rate
\begin{equation}\label{b3}
L_{\rm magn} = {2 \over {3 c^3}}\, B_{\rm s}^2 \sin^2\alpha \, R_{\rm s}^6 \, \Omega^4
\end{equation}
suggesting thus the following model of the neutron star spin-down:
\begin{equation}\label{b4}
L_{\rm sd} = L_{\rm magn}.
\end{equation}
The quantity $B_{\rm s} \sin\alpha$ can be inferred from $P$ and $\dot P$ for
a neutron star with known values of $I$ and $R_{\rm s}$.
For a large number of randomly oriented rotators the factor $\sin^2\alpha$ can be replaced
with its averaged value of $2/3$.

Another model, where the dipolar radiation is replaced with a magnetospheric wind of particles \cite{goldreich},
gives similar result as (\ref{b3}) for an orthogonal rotator:
\begin{equation}\label{b5}
L_{\rm sd} = L_{\rm wind} \approx {1 \over c^3}\, B_{\rm s}^2 \, R_{\rm s}^6 \, \Omega^4
\end{equation}
and therefore is independent of the angle $\alpha$.
Since there exists no observational support for $\dot P$ depending on $\sin\alpha$ the
standard approach is to apply the latter model to derive the strength of dipolar component of the magnetic field
\begin{equation}\label{b6}
B_{12}^2 =  10^{15}\, I_{45}\,R_6^{-6}\, P\, \dot P
\end{equation}
where $B_{12}\equiv B_{\rm s}/10^{12}\G$, and $R_6\equiv R_{\rm s}/10^{6}\cm$.

Assuming that $B_{\rm s}$ does not change with time one can integrate (\ref{b6}) to obtain 
the characteristic spin-down time scale $\tau$  -- a period of time elapsed since the pulsar
was born with initial period $P_{\rm i}$ 
\begin{equation}\label{b7}
\tau =  {P \over {2 \, \dot P}} \left[ 1 - {P_{\rm i}^2 \over P^2}\right] \s.
\end{equation}
As long as $P_{\rm i} \ll P$, which is thought to be satisfied for all classical pulsars and most
of millisecond pulsars, the last factor in (\ref{b7}) plays no role and thus
$\tau \approx P / 2\dot P$.

It is likely that neutron star magnetic fields contain high-order multipoles which
may dominate the dipolar component at the surface level. 
Their relative amplitudes as well as distribution remain, however, unknown.
It will be assumed throughout the paper that the dipolar magnetic field is not distorted by
rotational effects or  presence of strong outflowing wind of particles (the latter effect has been
recently invoked to decrease very high values of $B_{\rm s}$ inferred from $P$ and $\dot P$ for two SGRs \cite{kazanas};
in consequence, their classification as magnetars became questionable).
The field is therefore approximated with axisymmetric static dipole with field lines satisfying
$r \sin^{-2}\theta = R_{\rm dc}$ in polar coordinates $r$ and $\theta$, with the dipole constant $R_{\rm dc}$.
The dipole constant at which rigid rotation reaches speed-of-light limit is called the light cylinder: 
$R_{\rm lc} = c/\Omega$.
All field lines which cross the light cylinder are then considered as open lines, and their
footpoints on the stellar surface define two 
polar caps of radius $R_{\rm pc} \approx R_{\rm
s}\cdot (R_{\rm s}/R_{\rm lc})^{1/2}$, where the latter factor is the sine function of the polar
coordinate $\theta$ for the outer rim of the polar cap: $\sin \theta_{\rm pc} = (R_{\rm s}/R_{\rm lc})^{1/2}$.

\section{Observational overview of high energy domain}  
{\it A~posteriori} evidence that high-energy activity of pulsars must somehow draw from their rotational
energy $E_{\rm rot} = I {\Omega}^2/2$ comes from a simple finding that, essentially,
the success of detection of a particular pulsar in X-rays and/or gamma-rays 
was strongly correlated with its position in the lists of targets ranked by spin-down flux values $L_{\rm sd}/D^2$.

The aim of this section is to review recent observational status of RPP
in the high-energy (HE) domain, with X-rays included.
\begin{figure}
\begin{center}
\includegraphics[width=1.3\textwidth]{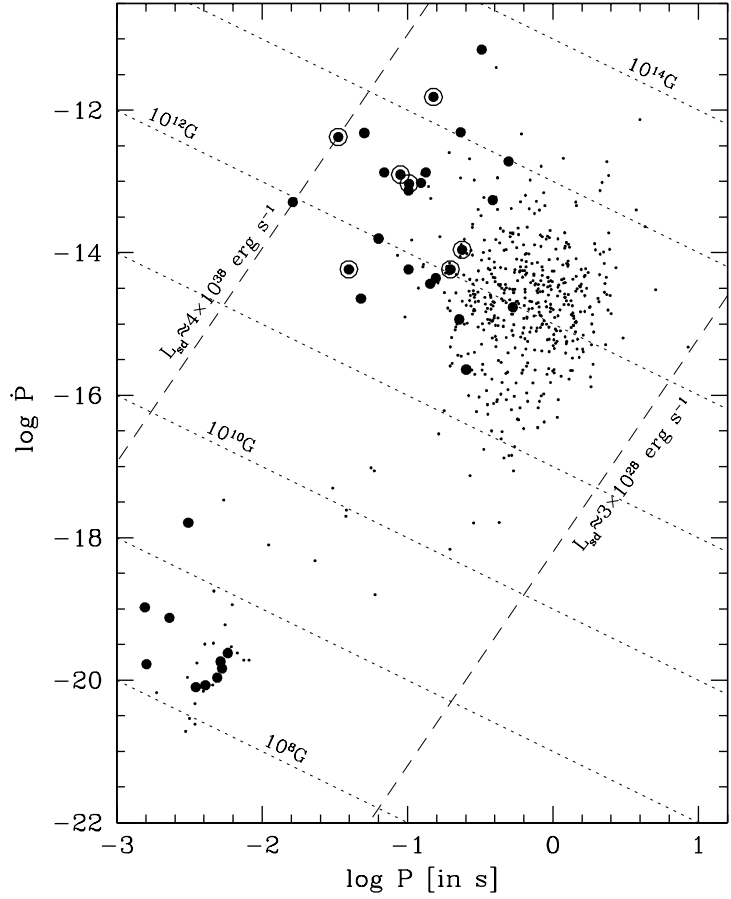}
\end{center}
\vspace*{-1.7cm}
\caption[]{$P-\dot P$ diagram for Rotation Powered Pulsars. The pulsars detected exclusively in radio
are indicated with dots; they are taken mostly from the data base of \cite{taylor}. 
Thirty five pulsars emitting X-rays are indicated with bullets.
These include two objects recently discovered with RXTE: J0537-6910  
in SNR N157B in LMC  \cite{marshall} is the fastest young pulsar known, spinning twice as fast as the Crab pulsar but with
similar value of spin down luminosity; J1846-0258 
in SNR Kes-75 \cite{gotthelf}, with $P = 0.32$s, the highest $\dot P$ 
among all RPP and no radio counterpart so far. Seven bullets in circles indicate
seven gamma-ray pulsars. Dashed lines correspond to constant values of the spin down luminosity $L_{\rm sd}$.
The upper line ($L_{\rm sd}\approx 4\times 10^{38}$erg~s$^{-1}$) includes the Crab pulsar and J1846-0258,
the lower one ($L_{\rm sd}\approx 3\times 10^{28}$erg~s$^{-1}$) includes J2144-3933 -- the slowest ($P = 8.5$s)
radio pulsar detected so far \cite{young}. Dotted lines correspond to constant values of the 
dipolar component of the surface magnetic field as inferred from $P$ and $\dot P$.
}
\label{ppdot}
\end{figure}
The HE domain is hereafter arbitrarily defined as extending from a fraction of $\keV$ up to about
$30\GeV$. Nevertheless, more emphasis is put on gamma-ray results.
Gamma-ray detections are particularly precious since their interpretation
is thought to be less ambiguous in comparison to X-ray detections. 
In the latter case (especially for very young objects) contributions
from initial cooling, internal friction, or other factors of {\it a~priori} unknown magnitude 
may dominate the X-ray emission (Note: indeed, four pulsars -- Vela, Geminga, B1055-52 and B0656+14 --
are classified as initial cooling candidates \cite{becker} since their X-ray emission is dominated by
a component which may be modelled by a blackbody emission from a NS surface).

For more than 1500 pulsars known to date only about 35 positive detections in X-rays and no more than
9 detections in gamma-rays have been achieved. 
There are firm detections by CGRO of 7 pulsars (dubbed {\it Seven Samurai})
and another 2 cases classified as `likely' detections. The gamma-ray sources
were identified by virtue of flux pulsations with previously known $P$ and $\dot P$. 
Crab and Vela are the only pulsars seen by all three instruments of CGRO.
No trace of pulsed signal in VHE range (300 GeV -- 30 TeV) has been found so far for 
the gamma-ray pulsars \cite{volk} \cite{sako} \cite{weekes}. 
However, strong steady VHE emission is associated with 3 out of 9 gamma-ray pulsars.
Two plerionic sources of the steady VHE radiation -- The Crab Nebula and the plerion around B1706-44 -- may serve as 
standard candles, with `grade A' according to \cite{weekes}. A third  plerion -- around the Vela pulsar -- was
given `grade B' in the same ranking.
All 9 gamma-ray pulsars are strong X-ray emitters. 

The positions of these HE pulsars are shown in the $P - \dot P$ diagram of Fig.\ref{ppdot} along with 
positions of about 700 radio pulsars for which $\dot P$ values were available. A remarkable fact is that the location
of X-ray sources does not correlate with the inferred  strength of magnetic field $B_{\rm s}$; at least not in 
a naively anticipated way that high-B objects would emit HE radiation, whereas low-B objects would not. In particular,
10 millisecond pulsars -- about thirty percent of all millisecond pulsars 
(the objects with $P \simless 0.01\s$ and $\dot P \simless 10^{-17}$, i.e. with low B values:
$B_{\rm s} \simless 10^9\G$) known to date -- have been detected as X-ray sources.
So far, millisecond pulsars eluded the detection in gamma rays and
just upper limits have been available for a handful of them from EGRET observations \cite{nel}.
In the case of J0437-4715 the upper limit is interestingly tight -- in disagreement
with the empirical relation $L_\gamma \propto L_{\rm sd}^{1/2}$ (see Fig.\ref{gamma}). Very recently, however,
the likely detection of pulsed gamma-ray emission from J0218+4232 has been reported \cite{kuiper_milli}.

\begin{figure} 
\begin{center}
\includegraphics[width=1.2\textwidth]{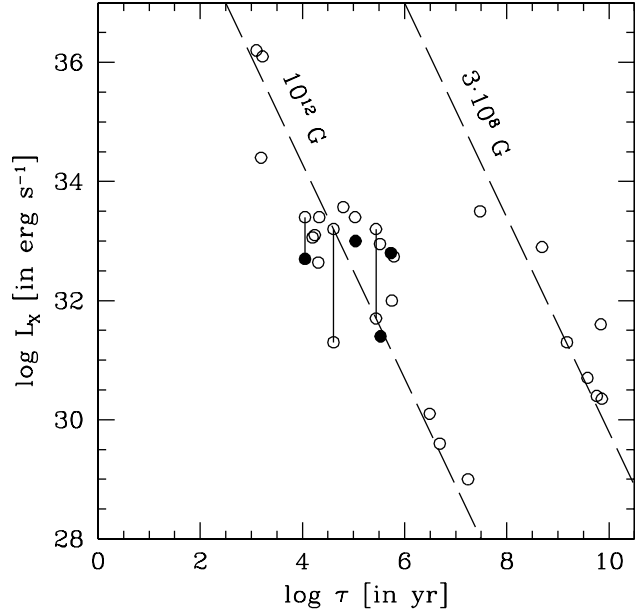}
\end{center}
\vspace*{-3.cm}
\caption[]{X-ray luminosity versus spin-down age $\tau$
for 29 out of 35 pulsars detected with ROSAT, ASCA and RXTE. Isotropic emission into $4\pi$ steradians
was assumed for pulsed and unpulsed sources to infer $L_{\rm X}$.
The empirical relation $L_{\rm X} \approx 0.001\,L_{\rm sd}$ found by \cite{becker} can be rewritten as
$L_{\rm X} \propto B_{\rm s}^{-2}\tau^{-2}$. If all classical and millisecond pulsars were to have
the surface magnetic field $B_{\rm s}$ of a fixed value $10^{12}$G and $3\times 10^{8}$G, respectively,
they would follow the two dashed lines labelled with $B_{\rm s}$.\\
Four filled circles are the initial cooling candidates; in increasing $\tau$ these are: B0833-45 (Vela) \cite{ogelman},
B0656+14 \cite{possenti}, J0633+17 (Geminga) \cite{halpern}, and B1055-52 \cite{slane}. 
In three cases the fitting of the data with either blackbody or power 
law spectral model was equally justified, but inferred X-ray luminosities are strongly model-dependent
(circles connected with vertical bars); in increasing $\tau$ these objects are: 
B0833-45 \cite{ogelman}, B2334+61, and B0114+58 \cite{lloyd}.
}
\label{coolx}
\end{figure}

\begin{figure}
\begin{center}
\includegraphics[width=0.8\textwidth]{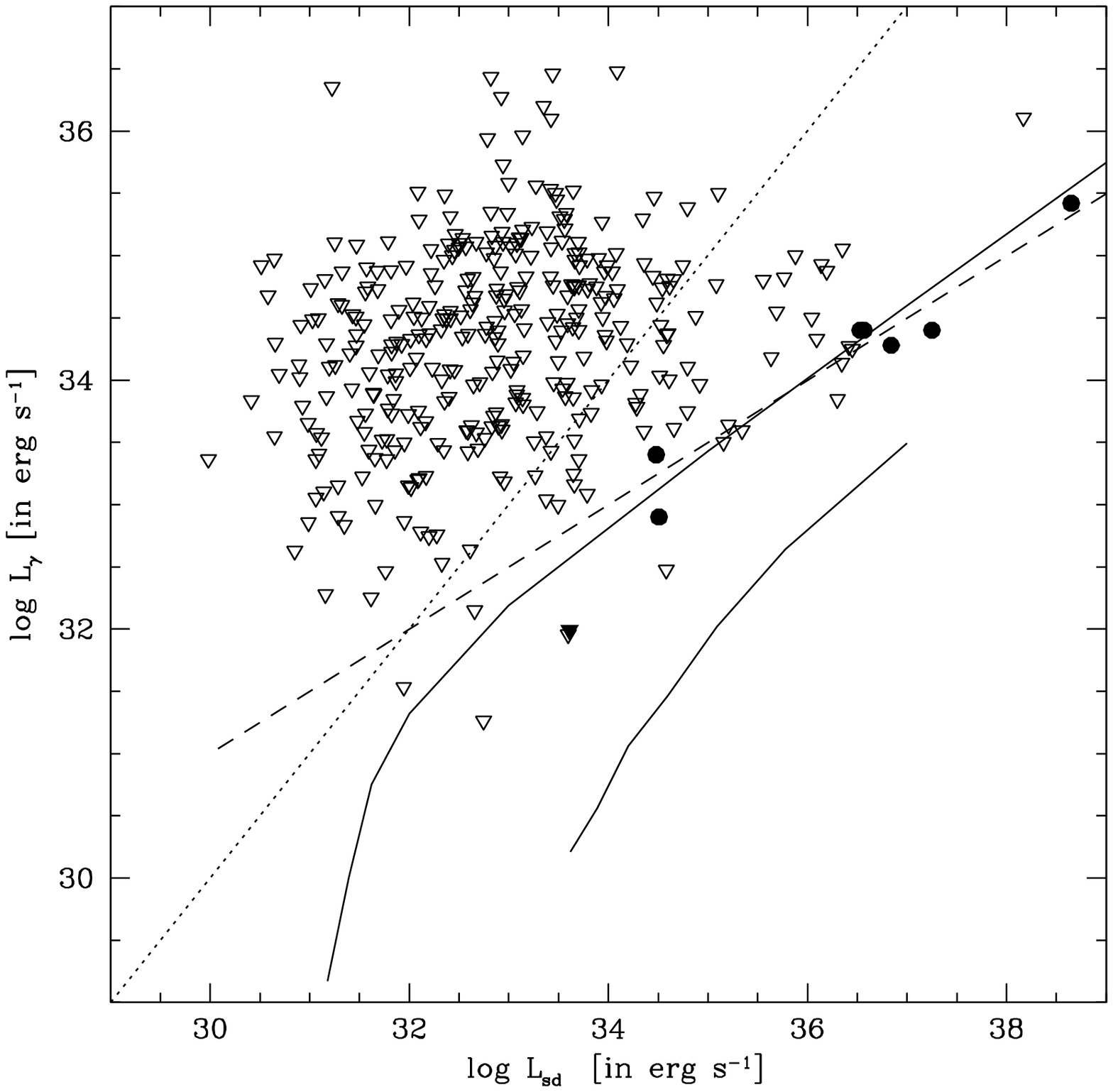}
\end{center}
\vspace*{-0.7cm}
\caption[]{Gamma-ray luminosity versus
spin-down luminosity for seven pulsars (filled dots) detected with the CGRO instruments.
Opening angle of one steradian was assumed for the gamma-ray emission.
Open triangles are the EGRET upper limits after \cite{nel} for 350 objects, including
seven millisecond pulsars.
The filled triangle indicates the upper limit for J0437-4715 \cite{fierro95}. Note, that most of the upper
limits are well above the maximum possible value for $L_\gamma$ set by $L_\gamma = L_{\rm sd}$ (dotted line).
The dashed line marks the empirical relation derived for the CGRO pulsars:
$L_\gamma \propto L_{\rm sd}^{1/2}$. Solid
lines show evolutionary tracks for a classical pulsar with $B_{\rm s} = 10^{12}$G (upper line)
and a millisecond pulsar with $B_{\rm s} = 10^9$G (lower line)
according to the phenomenological model of \cite{rudak98}. 
}
\label{gamma}
\end{figure}

Spectral analysis for pulsars detected with ROSAT PSPC ($0.1\keV$ to $2.4\keV$) shows that
in most cases a power-law spectral model provides acceptable fits to the data \cite{becker}.
Moreover, an intriguing empirical relation between inferred X-ray luminosity and spin down luminosity was found,
$L_X \approx 0.001\,L_{\rm sd}$, confirming rotational origin of most of the X-ray activity.
An interesting point is that the relation was obtained for all the sources regardless their 
temporal characteristics
(about 50~\% of all pulsars detected with ROSAT are unpulsed sources).
Fig.\ref{coolx} presents these results in a somewhat different way and for
slightly different values for $L_X$ (compiled by the author). 
A complementary empirical relation was found for pulsed emission from 19 pulsars observed 
with ASCA ($0.6\keV$ to $10\keV$).
Assuming opening angle of X-rays to be one steradian,
the inferred pulsed X-ray luminosity correlates with spin-down luminosity as 
\begin{equation}\label{Lx}
L_X =  10^{34}\left(\frac{L_{\rm sd}}{10^{38}\,\ergs}\right)^{3/2}\ergs,  
\end{equation}
according to \cite{saito}.

A similar power-law empirical relation holds for gamma-rays (cf. Fig.\ref{gamma}),
but with a different power-law index (e.g. \cite{thompson1055}):
\begin{equation}\label{Lg}
L_\gamma \simeq 10^{35}\left(\frac{L_{\rm sd}}{10^{38}\,\ergs}\right)^{1/2}\ergs.
\end{equation}

Important conclusion from Figs.\ref{coolx} and \ref{gamma} is that  neither $L_X$ nor $L_\gamma$
becomes a sizable fraction of $L_{\rm sd}$. The most efficient conversion 
of spin-down luminosity into high-energy radiation is taking place for 
B1055-52 -- the oldest pulsar among {\it Seven Samurai} -- with $L_\gamma \simeq 0.1 \, L_{\rm sd}$. 
Since in both, polar gap and outer gap models most of the energy gained by charged particles
in the gaps is transferred to HE photons, it means that the wind of particles leaving the magnetosphere
at the light cylinder is also unimportant energetically, i.e. $L_{\rm wind} \ll L_{\rm sd}$. 
Therefore, the lion's share of the spin down luminosity is probably carried away in a form of
magnetic dipole radiation $L_{\rm magn}$; in terms of the so called magnetization parameter $\sigma$
it means  $\sigma \gg 1$ at the light cylinder (cf. Sect.7).
 
\begin{figure}
\begin{center}
\includegraphics[width=1.1\textwidth]{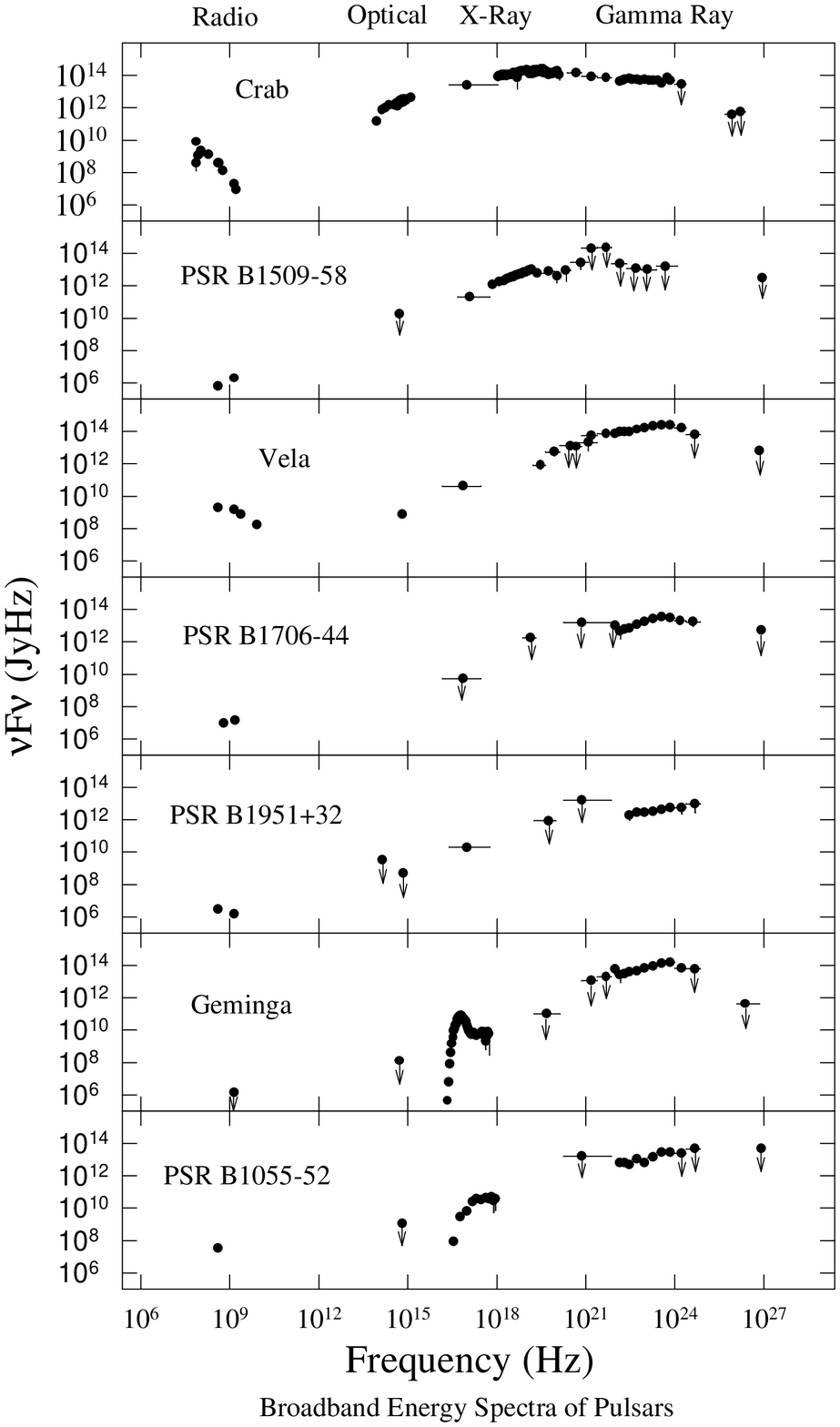}
\end{center}
\vspace*{-0.7cm}
\caption[]{Phase-averaged spectra for seven gamma-ray pulsars (Courtesy D.J.~Thompson).}
\label{nufnu}
\end{figure}

Broadband energy spectra per logarithmic
energy bandwidth extending from radio, optical and UV, to X-rays and gamma-rays, constructed for pulsed
phased-averaged components of {\it Seven Samurai} 
are particularly impressive \cite{thompson1055} and instructive. Fig.\ref{nufnu} reveals
substantial spectral differences among the objects, which became a subject of theoretical debates and speculations.

Short spectral characteristics of all 9 gamma-ray pulsars (including 2 `likely' sources) are given below
after \cite{thompson1055} for EGRET,
\cite{comptel_1st_cat} for COMPTEL and \cite{schroeder} for OSSE (and references therein). 
These instruments operating on board CGRO \cite{cgro_3} covered the following parts of HE domain: 
the Oriented Scintillation Spectrometer Experiment (OSSE) was operating in the energy range $50\keV - 10\MeV$, 
the Imaging Compton Telescope (COMPTEL) -- in the energy range $0.75\MeV - 30\MeV$,
and the Energetic Gamma Ray Experiment Telescope (EGRET) in the energy range $50\MeV - 30\GeV$. 

1) B0531+21 (Crab) -- Detected by EGRET,
COMPTEL, and OSSE. Its gamma-ray flux
consists of pulsed and unpulsed components,
the latter one coming from Crab Nebula. 
The overall phase-averaged photon spectrum 
in the range between 50 keV and 10 GeV is described satisfactorily
by a broken power-law shape 
${\rm d}N_{\gamma}/{\rm d}\varepsilon \propto \varepsilon^{-\alpha}$
with a break at $\varepsilon_{\rm br} \simeq 120 {~\rm{keV}}$, and
photon power-law index $\alpha = 1.71$ for $\varepsilon \leq \varepsilon_{\rm br}$, 
and $\alpha = 2.21$ for $\varepsilon > \varepsilon_{\rm br}$. The energy flux is
$f_\gamma \approx 7.3 \times 10^{-9}{{\rm erg}~{\rm s}^{-1} }{\rm{cm}^{-2} }$.

2) B1509-58 -- Detected by COMPTEL and OSSE. 
The initial COMPTEL detection was of marginal significance ($\sim 3\sigma$-detection)
in a narrow ($0.75 - 1\MeV$) energy band. However, recent analysis shows
that the spectrum extends to higher energies with a cutoff around $10\MeV$ \cite{kuiper} (Note: this new finding
is not marked in Fig.\ref{nufnu}).
The energy flux at 1 MeV may be as high as 
$1.4 \times 10^{-9}{~{\rm erg}~{\rm s}^{-1} }{\rm{cm}^{-2} }$, but the COMPTEL point
stands above the corresponding OSSE point by a factor of 4. The OSSE spectral fit
between 50 keV and $\sim 5 {~\rm{MeV}}$ with ${\rm d}N_{\gamma}/{\rm d}\varepsilon 
\propto \varepsilon^{-1.68}$ yields 
$f_\gamma \approx 5.6 \times 10^{-10}{~{\rm erg}~{\rm s}^{-1} }{\rm{cm}^{-2} }$.
EGRET put strong upper limits for photon flux above 100 MeV and 1 GeV , which clearly fall 
below simple power-law extrapolation of the OSSE spectral fit. That indicates a presence of spectral roll-over
at several MeV, in agreement with the cutoff claimed by  \cite{kuiper}.
The pulsar has the highest inferred magnetic field ($B_{\rm s} \simeq 1.5\times 10^{13}\G$ )
among seven gamma-ray pulsars, an essential point for explaining the 
cutoff at 10~MeV as due to photon-splitting effect \cite{splitting}.

3) B0833-45 (Vela) -- Detected by EGRET, COMPTEL, and OSSE. 
Its phase-averaged photon spectrum between 30~MeV
and 2~GeV can be reproduced as a power law with $\alpha = 1.7$, and a strong spectral break above $\sim 4\GeV$.
The spectrum flattens out in the OSSE range with $\alpha = 1.3$.
Estimated energy flux reaches $\sim 9 \times 10^{-9}{~{\rm erg}~{\rm s}^{-1} }{\rm{cm}^{-2} }$ (the brightest 
object in the gamma-ray sky).

4) B1706-44 -- Young Vela-like pulsar, detected by EGRET. 
The spectrum extends from $50\MeV$ beyond $10\GeV$ and may be
approximated with a broken power law, with photon index $\alpha$
changing from 1.27 to 2.25 at $1\GeV$.

5) B1951+32 -- Detected by EGRET and COMPTEL.
The spectrum extends from $0.75\MeV$ up to $30\GeV$ and may be
approximated with a single power law with photon index $\alpha = 1.89$.
Extremely sharp cut-off, with no apparent decline in the flux level.
But extrapolation towards TeV falls 2 order of magnitude above
an upper limit (not shown in Fig.\ref{nufnu}) set by Whipple group.

6) J0633+17 (Geminga) -- Confirmed detection by EGRET only.
The photon spectrum may be approximated with a single power law, with $\alpha = 1.50$, 
extending from $30\MeV$ to a roll-off at $2\GeV$.

7) B1055-52 -- Detected by EGRET above 70 MeV. Its spectrum can be represented
by a single power law with photon index $\alpha = 1.73$,
and a possible break around $1\GeV$.

8) B0656+14 -- A $3\sigma$-detection by EGRET had been reported. The pulsar with its parameters 
($P$, $\dot P$, and particularly $L_{\rm sd}$) resembles Geminga nad B1055-52.
Its photon spectrum estimated for low number of events may be
represented  between $10\MeV $and $10\GeV$ as a very steep power-law with
the index $\alpha = 2.8$.

9) J0218+4232 -- Marginal detection by EGRET (at $3.5\,\sigma$ level) of pulsed emission 
has been reported recently in the energy range $100\MeV - 300\MeV$ for this distant ($D \simgreat 5.85\,$kpc)
millisecond  ($P =0.0023\s$) pulsar in a binary system with a low-mass white dwarf \cite{kuiper_milli}.  
The inferred luminosity of the pulsed emission for 1 steradian opening angle
reaches $L_\gamma \simeq 1.64\times 10^{34}\ergs \simeq 0.07\, L_{\rm sd}$.

Gamma-ray light curves differ significantly from those  in X-rays, optical, and radio.  
Their most striking feature are relatively long duty cycles as well as phase shifts in comparison to the radio pulses.
Only for the Crab pulsar the peaks in gamma-rays as well as in radio wavelengths occur at the same
rotational phases.
The light-curve shapes fall into two categories. The Crab pulsar, Vela and Geminga
show two sharp pulses separated in phase by $0.4 - 0.5$ and connected
by an interpulse bridge of considerable level.
B1706-44 shows two peaks separated by 0.2 in phase, with some hints
of a third component in between. Other pulsars exhibit broad single pulses. 
Unknown opening angles for gamma-ray emission introduce a factor of uncertainty when inferring 
the gamma-ray luminosities. Broad peaks in gamma-ray pulses do not necessarily mean
large opening angles for gamma-ray emission. Polar cap
models, which rely on purely dipolar magnetic fields postulate nearly aligned rotators,
where inclination of magnetic axis to spin axis is  
comparable to the angular extent of the polar cap \cite{obliquity}.

With a dicovery of sharply peaked pulsed X-ray emission in the fastest millisecond pulsar B1937+21
an apparently separate group of millisecond X-ray pulsars emerges,
with its members -- B1821-24 \cite{saito}, J0218+4232 \cite{kuiper_milli}, 
and B1937+21 \cite{takahashi} -- being scaled-down versions of the Crab 
pulsar as far as sharp pulse profiles and
hard power-law X-ray spectra are concerned. An astonishing common feature within the group is the same strength
of  the magnetic field estimated at the light cylinder and the fact that it matches the strength of the Crab pulsar 
magnetic field at the light cylinder.

\section{Unipolar induction -- a toy model}
\subsection{Vacuum rotator}
Let us begin with a frequently invoked order-of-magnitude estimate advertising
rotating neutron stars as potentially powerful accelerators and thus good candidates to
explain UHECR (the problem addressed in Sect.7).
Consider a neutron star of radius $R_{\rm s}$ and surface magnetic field $B_{\rm s}$
as a perfect conductor.
For the rotating star with its dipolar magnetic field immersed in a vacuum 
an external quadrupole electric field $\vec {\cal E}$ developes, with non-zero
component along magnetic field lines at the surface.
The corresponding electrostatic potential $\Phi$  in polar coordinates $r$ and $\theta$ reads \cite{michel}
\begin{equation}\label{toy1}
\Phi (r,\theta) = - {Q \over r} \, \left(3 \cos^2\theta - 1 \right),
\end{equation}
where $Q \propto B_{\rm s}/P$ is the quadrupole moment.
Maximal electromotive force will then be induced between one of the two poles of the star and its equator: 
\begin{equation}\label{toy2}
\Delta\Phi_{\rm equator} = \frac{1}{2c}\, B\, \Omega
\, R^2_{\rm s}. 
\end{equation}
The corresponding voltage drop 
\begin{equation}\label{toy3}
\Delta V_{\rm equator} \approx
3 \times 10^{16} B_{12} R_6^2 P^{-1} {\rm Volt}
\end{equation}
reaches huge values. If such a unipolar inductor was to operate in the Crab pulsar, B1509 or J1846-0258 
(see Fig.\ref{ppdot}), it
would bring a fully ionized atom of iron ($Z = 26$) close to the energy of $10^{20}$eV.
However, the assumption about a vacuum surrounding the entire star is not correct. 
The space containing field lines closed within the light cylinder is expected to fill in quickly with
trapped charged particles supporting the electric field which forces them to corotate with the star 
(cf. the subsection below).   
Therefore, the only regions on the stellar surface appropriate for
the unipolar induction to act are those containing open field lines only, i.e. the polar caps.
From (\ref{toy1}) and  (\ref{toy2}) the potential difference between the pole and the outer rim of the polar cap
follows as 
\begin{equation}\label{toy4}
\Delta\Phi_{\rm pc} = \Delta\Phi_{\rm equator} \left(\frac{R_{\rm pc}}{R_{\rm s}}\right)^2,
\end{equation}
with voltage drop of
\begin{equation}\label{toy5}
\Delta V_{\rm pc} \approx 7 \times 10^{12}\, B_{12}\, P^{-2}\, {\rm Volt}.
\end{equation}
\subsection{Rotating magnetosphere}
We will concentrate hereafter on a class of models where the supply of
charged particles from a neutron star surface along open filed lines 
is not limited by binding or cohesive energy of the particles
and therefore can reach the so-called Goldreich--Julian rate at the surface.  
Such a supply of charges was dubbed `Space Charge Limited Flow'
\cite{sturrock}, \cite{arons79} or `free emission'. 
A concise but to-the-point account of essential properties of the SCLF models 
has been presented recently by \cite{arons00}.

Three boundary conditions essential for the electrodynamics above the polar cap are \cite{tsygan}:\\
1) $\vec {\cal E} \cdot \vec B = 0$ for the magnetosphere within the closed field lines,\\
2) $\Phi = 0$ at the surface and at the interface between the closed
magnetosphere and the open field lines,\\
3) ${\cal E}_\parallel = 0$ at the surface level.\\
Last but not least, it is assumed that the outflow is stationary and the magnetosphere remains axisymmetric.

The electric field $\vec {\cal E}$ required to bring a charged particle into corotation is
\begin{equation}\label{1}
  \vec {\cal E} + \frac{1}{c} ((\vec \Omega - \vec \omega_{\rm LT})
   \times \vec r) \times \vec B = 0
\end{equation}
where the inertial-frame dragging effect is included \cite{tsygan} with  
$\vec \omega_{\rm LT} = \kappa_{\rm g} (R_{\rm s}/r)^3\,\vec \Omega$,
and $\kappa_{\rm g} = 2GI/R_{\rm s}^3 c^2$.

Charge density necessary to support this local $\vec {\cal E}$ is
\begin{equation}\label{2}
\varrho_{corot} = \frac{1}{4\pi}
\vec \nabla  \cdot \vec {\cal E} \approx -\frac{\vec \Omega \cdot \vec B}{2\pi c}\, \left[1 - \kappa_{\rm g}
\left(\frac{R_{\rm s}}{r}\right)^3\right]
\end{equation}
The charge density $\varrho_{corot}$ due to SCLF at $r = R_{\rm s}$ is called Goldreich--Julian charge density
and it is labelled with `GJ':
\begin{equation}\label{4}
\varrho_{\rm GJ} \approx -\frac{\vec \Omega \cdot \vec B}{2\pi c}\,\, [1 - \kappa_{\rm g}].
\end{equation}
As charged particles flow out along the open field lines a deviation of the local
charge density $\varrho_{\rm local}$  from the local corotation density $\varrho_{corot}$ develops.
By using now two relations satisfied 
in the dipolar structure, $B(r) \propto r^{-3}$ and $\varrho (r) \propto r^{-3}$, one obtains 
a simple formula for local deviation from the corotation charge density:
\begin{equation}\label{5}
\varrho_{\rm local} - \varrho_{corot} = \frac{\vec \Omega \cdot \vec B}{2\pi c}\,\, \kappa_{\rm g}\,\, \left[1 -
\left(\frac{R_{\rm s}}{r}\right)^3\right].
\end{equation}
Accordingly, the accelerating potential drop in SCLF reads
\begin{equation}\label{3}
\Delta\Phi_\| \approx \Delta\Phi_{\rm pc}\, \,\kappa_{\rm g} \, \left[1 -
\left(\frac{R_{\rm s}}{r}\right)^3\right].
\end{equation}
This is a remarkable result obtained by \cite{tsygan}: due to the inertial-frame dragging effect the 
particles drop through the potential which is significantly larger than in the classical approach of \cite{arons79}
(for $P = 1\s$ it becomes $10$ times larger).
Moreover, the electric field ${\cal E}_\parallel$ developes now along all open lines regardless their orientation
with respect to the spin axis (the effect is not shown in this simplified presentation). 
Therefore, all open field lines are `favourable', in contrast to \cite{arons79}. 

\section{Electric field structure in SCLF gaps}
The model with frame dragging effects \cite{tsygan}, presented in a simplified form
in subsection 4.2, does not take into account possible feed-back effect due to
$e^\pm$-pairs formed via photon absorption within open magnetic field lines. Copious pair formation occurs 
in a relatively thin
layer called for this reason a pair formation front (PFF). The creation of pairs leads 
to screening of
the accelerating field ${\cal E}_\|$ within the layer of PFF.
A detailed picture of this effect would require to follow the dynamics of electrons and positrons
in a self-consistent way. Instead, it is reasonable to assume, that  the field is shorted out
at the height were the first $e^\pm$-pair is created (hereafter denoted as $h_c$): ${\cal E}_\| = 0$ for $h \ge h_c$.

The problem of electric field structure in the context of SCLF with  boundary condition ${\cal E}_\| = 0$ set at
$h_0 = 0$ (stellar surface) and at $h \ge h_c$ (PFF)
was formulated and solved by \cite{muslimov}.
The solution is rather lengthy and includes special functions. It is however possible to
obtain simple but quite accurate analytical approximations. 
As long as the length $h_c$ of the accelerator is of the order of polar cap radius
$R_{\rm pc}$, the accelerating electric field 
may be approximated according to \cite{dyks_acc} as
\begin{equation}
{\cal E}_\parallel \approx -1.46\ \frac{B_{12}}{P^{3/2}}\ h \left(1 - \frac{h}{h_c}\right) 
f_1(\xi) \cos\chi\ {\rm Gauss},
\label{electric1}
\end{equation}
where
$B_{12}=B_{\rm s}/10^{12}\G$, $P$ is the spin period in seconds, $\chi$ is the angle between
the spin axis and the magnetic moment of the rotating star, $h$
is expressed in cm, and $M=1.4\, M_\odot$, $R_{\rm s}=10^6$cm.
The magnetic colatitude $\xi\equiv \theta/\theta(\eta)$ 
is scaled with the half-opening angle of the polar magnetic flux tube $\theta(\eta)$, 
where $\eta\equiv 1 + h/R_{\rm s}$. The magnetic colatitude function $f_1(\xi)$ is a monotonically
decreasing function, with $f_1(0) \simeq 1$ and  $f_1(1) = 0$.

Vertical structure of the electric field depends (via the location of PFF) on radiative processes
which induce the pair creation: curvature radiation and
inverse Compton scattering on soft X-ray photons from the stellar surface (brief characteristics 
of these processes is presented in the next section). 
An interesting effect was noticed in this context \cite{muslimov}: Suppose that a small fraction of positrons is 
stopped by a residual (non-zero) electric
field at the site of their creation and then forced to flow towards the stellar surface (the effect
noticed already by \cite{sturrock}). The backflowing positrons are expected to induce a formation 
of an additional PFF, which would short out the electric field at $h_0$.
These positrons cool upon the
action of ICS and CR. With reasonable surface temperatures ($T_{\rm s} \sim 5\times 10^5\K$) 
it is ICS which dominates the cooling
of  upward moving electrons and downward moving positrons. Therefore, the $e^\pm$-pair creation
will be induced by upscattered photons from the stellar surface, rather than by curvature photons. 
The situation is not symmetrical, however, for  electrons and positrons since the field of soft photons
is not symmetrical with respect to both types of particles. In consequence, the positrons cool
more efficiently, the ICS-induced cascades are easier to achieve for positrons than electrons and thus
a lower PFF (where ${\cal E}_\| = 0$) tends to be located above the stellar surface. 
Such a situation is not stable, therefore.
However, elevating the accelerator up to altitude $h_0 \sim 1\, R_{\rm s}$ above the surface diminishes the role
of the ICS; the CR cooling dominates here and a stable accelerator is possible.
No self-consistent calculations of such a `sandwich-like' accelerator exist at present, but 
the results obtained by \cite{muslimov} with an approximate treatment of the problem
look promising indeed.


\section{Radiative processes in pulsar magnetospheres}
Cooling of ultrarelativistic electrons via curvature radiation (CR) and magnetic inverse Compton scattering (ICS) 
are the most natural ways of producing hard gamma-rays
 capable of inducing
cascades of $e^\pm$-pairs and secondary HE photons.
These two processes dominate within two distinct ranges of Lorentz
 factors $\gamma$ of primary electrons.

When $\gamma \simless 10^6$,
 magnetic inverse Compton scattering  plays a dominant role in braking the electrons 
and it is the main source of hard
 gamma-ray photons \cite{sdm}. 
Energy losses due to resonant ICS
 limit the Lorentz factors of the particles to a level 
which depends on electric field strength $\cal E_\parallel$, temperature $T$ and size of 
hot polar cap, and magnetic field strength $B_{\rm s}$ \cite{xia}\cite{sturner}.
The Lorentz factors
 can then be limited even to $\sim 10^3$.
 This stopping effect becomes more efficient for stronger
magnetic fields, 
and it was suggested as an explanation for 
the observed cutoff at $\sim 10\MeV$ in the spectrum of B1509-58 \cite{sturner}.

However, in their modern versions the accelerators of particles are strong enough to outpower the ICS cooling.
In consequence, very high Lorentz factors -- $\gamma\simgreat 10^6$ -- are achieved by electrons, limited by CR.  
The first detailed scenario of radiative processes in CR-induced cascades was presented by \cite{dau82} and
despite many modifications and additions its basic features remain valid. The model assumes
that primary electrons accelerated to ultrarelativistic energies emit curvature photons which in turn are
absorbed by the magnetic field and $e^\pm$-pairs are created. These pairs cool off instantly
via synchrotron radiation process. Whenever the SR photons are energetic enough they may
lead to further creation of pairs, etc..
ICS can still be incorporated to the models with CR-induced cascades as the process relevant for  
$e^\pm$-pairs, since typical Lorentz factors of theirs do not exceed $\sim 10^3$. 
According to the analytical model of \cite{bing} the empirical relations
for X-ray and gamma-ray luminosities of pulsars (presented in Sect.3) can be reproduced satisfactorily 
when the ICS involving $e^\pm$-pairs is included.

Processes relevant for production and transfer of HE radiation in pulsar magnetospheres are, therefore: \\
 - Curvature Radiation,\\
 - Magnetic Inverse Compton Scattering, \\
 - Magnetic Pair Creation \quad $\gamma \, \rightarrow e^\pm$ \\
 - Synchrotron Radiation,\\
 - Photon splitting \quad $\gamma \rightarrow \gamma + \, \gamma$,\\
 - Photon-photon Pair Creation \quad $\gamma + \, \gamma \rightarrow e^\pm$.
 
Basic properties of these processes are briefly reviewed below. The last process in the list has been omitted.
The reason is that, whenever recalled in the context of pulsar magnetospheres, 
photon-photon pair creation is treated exactly as in free space. 
Such treatment is justified in models of `thick outer gaps' \cite{cheng} but within the framework
of polar cap models this is not the case, in general. 
However, no handy formula is available for the cross section of this process in the limit of high $B$ 
and standard non-magnetic formulae are in use whenever necessary.

In order to illustrate the significance of these processes in forming HE spectrum extending over
many decades in energy, the numerically calculated effects due to the first 4 processes in the list 
will be presented in Fig.\ref{vela} (after \cite{dyks}) along with overlaid data points for
the Vela pulsar. The dipolar field in the Vela pulsar does not exceed $10^{13}\G$ 
and thus photon splitting is not competitive to magnetic pair creation; its effects are negligible and will not shown. 
The electric field structure of the accelerator used in these calculations is taken after \cite{muslimov}
and appropriate rescaling due to $h_0 > 0$. However, the case calculated for Fig.\ref{vela} was chosen with
$h_0 = 3 R_{\rm s}$, i.e. much higher than in \cite{muslimov}, in order
to better reproduce the phase averaged spectrum of the Vela pulsar.

\subsection{Curvature radiation}
Relativistic electron of energy $\gamma m c^2$ (we take $\gamma \gg 1$) 
sliding along the magnetic field line
of curvature $\varrho_{\rm cr}$ will emit photons with a continuum  energy spectrum 
peaked at
\begin{equation}
\varepsilon_{\rm peak} \approx 0.29\, \varepsilon_{\rm cr}\, ,
\label{cr0}
\end{equation}
where   
\begin{equation}
\varepsilon_{\rm cr} = {3 \over 2} \, c \, \hbar \, {\gamma^3 \over \varrho_{\rm cr}},
\label{cr1}
\end{equation}
is called the characteristic energy of CR. The radius $\varrho_{\rm cr}$
for a purely dipolar line attached to the outer rim of the polar cap can be approximated
not far away from the NS surface as  $\varrho_{\rm cr} \approx \sqrt{ R_{\rm s} \cdot R_{\rm lc}} \approx 10^8 \sqrt{P} \cm$.
The cooling rate of that electron is
\begin{equation}
\dot \gamma_{\rm cr} = - {2 \over 3} \, {e^2 \over m c}\, {\gamma^4 \over \varrho_{\rm cr}^2}.
\label{cr10}
\end{equation}
For a monoenergetic injection function of electrons $Q(\gamma) \propto \delta (\gamma - \gamma_0)$
and their cooling due solely to CR the electrons will assume a single power-law distribution
in energy space $N_{\gamma}({\rm el.}) \propto \gamma^{-4}$ for $\gamma < \gamma_0$ as long as they stay
within the region of the cooling. Their escape introduces a natural low-energy 
cutoff $\gamma_{\rm cutoff}$ in $N_{\gamma}({\rm el.})$.
Therefore, the unabsorbed CR energy spectrum $f_\varepsilon (\varepsilon)$ due to the injected 
electrons has a broken power-law shape, with a high-energy limit set by $\gamma_0$
and the break at some energy $\varepsilon_{\rm break}$.
For $\varepsilon > \varepsilon_{\rm break}$ the energy spectrum is
$f_\varepsilon (\varepsilon) \propto \varepsilon^{-2/3}$, and 
$f_\varepsilon (\varepsilon) \propto \varepsilon^{+1/3}$ for $\varepsilon < \varepsilon_{\rm break}$.
Since nonthermal spectra cover ususally many decades in energy it is more
convenient to use $\varepsilon f_\varepsilon (\varepsilon)$ 
for easy comparison of power in different parts of energy space
(see Figs.\ref{nufnu} and \ref{vela}).
Accordingly,
$\varepsilon f_\varepsilon (\varepsilon) \propto \varepsilon^{+1/3}$ above the break, and  $\propto \varepsilon^{+4/3}$
below the break.

The cutoff limit $\gamma_{\rm cutoff}$ can be found by comparing the characteristic cooling time scale
$t_{\rm cr} \equiv \gamma/ |\dot \gamma_{\rm cr}|$ with the estimated time of escape  $t_{\rm esc}$,
which we take as  $t_{\rm esc} \approx \varrho_{\rm cr}/ c$. 
Therefore 
\begin{equation}
\varepsilon_{\rm break} \approx 0.29 \cdot {9 \over 4} \, \hbar \, {c \over r_0} \approx 44 \MeV,
\label{cr3}
\end{equation}
where $r_0$ is the classical electron radius. Note, that 
the photon energy $\varepsilon_{\rm break}$ at which the spectral break occurs does not depend on any
pulsar parameters. 

The spectrum of CR calculated  numerically to model the Vela pulsar \cite{dyks}
is shown in Fig.\ref{vela} as dot-dashed line. 
High-energy cutoff due to one-photon magnetic absorption occurs around $10\,$GeV. Note the importance of 
gamma-ray detectors capable to operate above $10\,$GeV for (in)validating the model.
The low-energy CR spectral break $\varepsilon_{\rm break}$ is prominent at $\sim 40 \MeV$.
Below $\varepsilon_{\rm break}$ the power of CR decreases and eventually  
becomes unimportant at $\sim 1 \MeV$ where the synchrotron component takes over.

\subsection{Magnetic pair creation}
Pair creation via magnetic photon absorption ($\gamma + \vec B
\rightarrow e^\pm + \vec B$) is kinematically correct since the magnetic field
can absorb momentum. To ensure high chances for the process to occur it is not
enough for a photon propagating at some angle $\alpha$ to local $\vec B$
to satisfy 
the energy threshold condition, $\sin \psi \cdot \varepsilon \ge 2 m c^2$,
but high optical thickness $\tau_{\gamma B}$ within the magnetosphere is required.
In fact the condition $\tau_{\gamma B} = 1$ has been used as a criterium for the so called death-line 
for radiopulsars in the $P - \dot P$ diagram. Maximal values of $\sin \psi$ for 
curvature photons in the dipolar field do not exceed
$\sim 0.1 \sin \theta_{\rm pc} \approx 0.0014\, P^{-1/2}$ \cite{sturrock},
so the both conditions are difficult to meet for long-period rotators.

The absorption coefficient for the process as
described in \cite{erber} and used to calculate $\tau_{\gamma B}$  reads 
\begin{equation}
\eta(\varepsilon) = {1\over 2} {\alpha \over \lambdabar_{\rm c}} {B_\perp
\over B_{\rm crit}} T\left(\chi \right)
\label{mpp1}
\end {equation}
where $\alpha$ is a fine structure constant, $\lambdabar_{\rm c}$
is a Compton wavelength, 
$B_{\rm crit} = m^2 c^3/e\hbar \simeq 4.4\times 10^{13}\G$,
$B_\perp$ is the component of the
magnetic field perpendicular to the photon momentum, and $\chi
\equiv {1\over 2}{B_\perp \over B_{\rm crit} }{\varepsilon\over
m_e c^2}$ is the Erber parameter $\chi$.  The function $T(\chi)$ is then approximated as
 $T(\chi) \approx 0.46 \exp\left(-4 f/3\chi
\right)$, 
which is valid for $\chi \simless 0.2$; for $\chi \simgreat 0.2$
this approximation starts to overestimate $\eta$.
The function $f$ is the near-threshold correction introduced
by \cite{dau83} important in particular in the case of classical pulsars. 
Electron-positron pairs created through the magnetic absorption
radiate then via the synchrotron process (SR) described below.

\subsection{Synchrotron radiation}
Consider a particle of energy $\gamma m c^2$ gyrating around a local field line at a pitch angle $\psi$.
Let $\gamma_\parallel$ denotes the Lorentz factor of the reference frame comoving with the center
of the gyration. As long as $\gamma_\parallel \gg 1$  it relates to the pitch angle $\psi$
via $\sin \psi \approx \gamma_\parallel^{-1}$.
The energy available for synchrotron emission at the expense of the particle is  $\gamma_\perp m c^2$,
and $\gamma = \gamma_\perp \, \gamma_\parallel$. 

The rate of SR cooling reads
\begin{equation}
\dot \gamma_{\rm sr} = - {2 \over 3} \, 
{r_0^2 \over m_{\rm e} c}\, B^2 \gamma_\perp^2 = - {2 \over 3} \,
{r_0^2 \over m_{\rm e} c}\, B^2 \sin ^2\psi \,\gamma^2.
\label{sr0}
\end{equation}
In comparison to the CR cooling it is  enormous (due to much smaller curvature radius). 

Critical photon energy (analogous to (\ref{cr10})) reads 
\begin{equation}
\varepsilon_{\rm sr} = {3 \over 2} \, \hbar \, {e B \over m_{\rm e} c} \gamma^2 \sin \psi.
\label{sr1}
\end{equation}

For a monoenergetic injection function of particles ($e^\pm$-pairs in the context of this review) 
and their cooling due to SR the energy spectrum of SR  
spreads between a high-energy limit $\varepsilon_{\rm sr}(\gamma_0)$ set by 
$\gamma_0$ of the injected (created) particles, and 
a low-energy turnover $\varepsilon_{\rm ct}$ determined by the condition $\gamma_\perp \sim 1$:
\begin{equation}
\varepsilon_{\rm ct} \equiv \varepsilon_{\rm sr}(\gamma = \gamma_\parallel) 
= {3 \over 2} \, \hbar \, {e B \over m_{\rm e} c} \,{1 \over {\sin \psi}}.
\label{sr2}
\end{equation}
The spectrum assumes a single power-law shape $f_\varepsilon (\varepsilon) \propto \varepsilon^{-1/2}$
(and accordingly -- 
$\varepsilon f_\varepsilon (\varepsilon) \propto \varepsilon^{+1/2}$) above the turnover.
Below $\varepsilon_{\rm ct}$, the spectrum $f_\varepsilon$
changes it slope, asymptotically reaching $\propto \varepsilon^{+2}$.
It is built up by contributions from low-energy tails emitted by particles with $\gamma_\perp \gg 1$, 
and each low-energy tail is assumed to cut off at local gyrofrequency, which in the reference frame
comoving with the center of gyration is $\omega_B = {e B \over m_{\rm e} c \,\gamma_\perp}$.
 
\begin{figure}
\begin{center}
\includegraphics[width=1.0\textwidth]{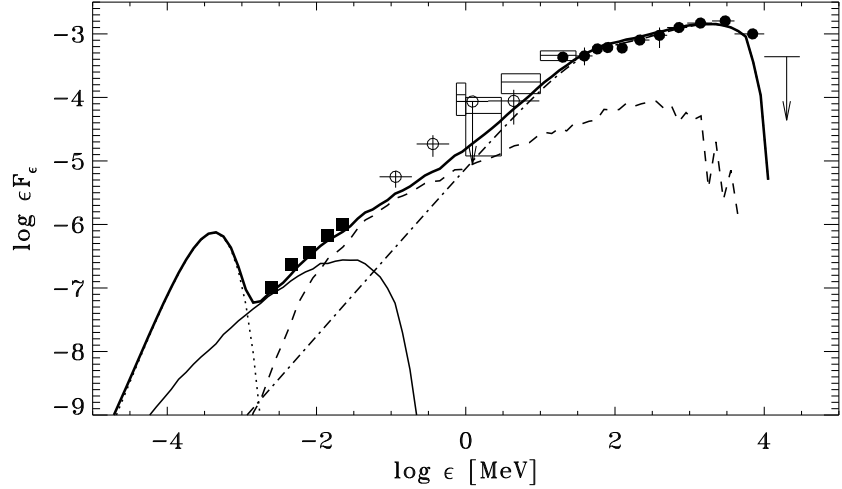}
\end{center}
\vspace*{-0.7cm}
\caption[]{The model energy spectrum calculated by \cite{dyks} to reproduce the spectral 
features of the Vela pulsar ($P=89$ ms, $B_{\rm s} = 6\times 10^{12}$G).
The accelerator is located at $h_0 = 3 R_{\rm s}$ above the surface (see Sect.6).
The broad--band spectrum consists of four components due to:
curvature radiation of primary electrons (dot-dashed), synchrotron radiation of
secondary $e^\pm$--pairs (dashed), inverse Compton scattering of surface X-ray photons on
the $e^\pm$--pairs (thin solid) and the blackbody surface emission (dotted). 
The surface temperature $T_{\rm s} = 1.26\times 10^6$K was assumed.
Total spectrum  is given by a thick solid line.
Phase-averaged data points for Vela from different satellite experiments are indicated.
Filled squares -- RXTE \cite{vela_rxte}; open circles -- OSSE \cite{vela_osse};
open squares -- COMPTEL \cite{comptel_1st_cat}; filled circles plus upper 
limit just above $10$~GeV -- EGRET \cite{cgro}. Vertical axis is in log of MeV~cm$^{-2}$~s$^{-1}$ units.
}
\label{vela}
\end{figure}

The spectrum of SR calculated with Monte-Carlo method to model the Vela pulsar is shown in
Fig.\ref{vela} as a dashed line.  
The low-energy part of the SR spectrum at $\varepsilon_{\rm ct}$ seems to be essential for 
reproducing an interpolation between the RXTE and the OSSE data.
\subsection{Magnetic Inverse Compton Scattering}
Consider an electron with a Lorentz factor $\gamma$ moving along a magnetic field line $\vec B$
and a photon of energy $\varepsilon = \epsilon m c^2$ moving at angle $\arccos{\mu}$ to the
field line. 
In the reference frame comoving with the electron (primed symbols)
the counterpart of the free-space Compton formula, due to energy-momentum conservation
appropriate for collisions with the electron at the ground Landau
level both in the initial and final state, reads 
\begin{eqnarray}
\lefteqn{ \epr_s = \left(1 -
{\mu^\prime_s}^2\right)^{-1}\left\{\vbox{\vskip5mm} 
1 + \epr(1 - \mu^\prime\mu^\prime_s)\ + \right.}\nonumber\\
&&\hbox{\hskip4mm}\left.-\left[  1 + 2\epr\mu^\prime_s (\mu^\prime_s -
\mu^\prime) +
{\epr}^2 (\mu^\prime_s - \mu^\prime)^2  \right]^{1/2}\right\}
\end{eqnarray}
where $\epr=\epsilon\gamma(1 - \beta\mu)$ \cite{herold}, and
symbols with no subscript and  with the subscript `s' refer to the state before the scattering and 
after the scattering, respectively.
A longitudinal momentum of the electron 
in the electron rest frame changes due to recoil from zero to $(\epr\mu^\prime - \epr_s\mu^\prime_s)mc$.

The polarization-averaged relativistic magnetic
cross section in the Thomson regime may be approximated 
with a nonrelativistic formula \cite{dermer}:
\begin{equation}
\sigma = \frac{\st}{2}\left(1 - {\mu^\prime}^2 + (1 +{\mu^\prime}^2)
\left[g_1 + \frac{g_2 - g_1}{2}\right]\right)
\label{crosssection}
\end{equation}  
where $\st$ is the Thomson cross section, and $g_1$ and $g_2$ are given by
\begin{equation}
g_1(u) = \frac{u^2}{(u+1)^2}, \hbox{\hskip 1cm}
g_2(u) = \frac{u^2}{(u-1)^2 + a^2}
\end{equation}
where $u\equiv\epr/\eb$, $a\equiv 2 \alpha\eb/3$, $\eb \equiv \hbar e B/m^2c^3$ and $\alpha$ is a
fine-structure constant. 
The resonance condition for the scattering is therefore the cyclotron resonance $\epr = \eb$.
The factor $a$ represents a `natural' broadening of the resonance due to finite lifetime at the excited 
Landau level.

In the Klein-Nishina regime ($\epr > 1$) the relativistic magnetic cross 
section for the $|\mu^\prime|\approx 1$ case becomes better approximated with the well
known Klein-Nishina relativistic nonmagnetic total cross section $\skn$
\cite{dau86} \cite{dermer}.

The rate $\cal R$ of scatterings subject by an electron moving across the field of soft photons,
measured in the lab frame is
\begin{equation}
{\cal R} = c \int d\Omega \int d\eps\ \sigma
\left(\frac{dn_{\rm ph}}{d\eps d\Omega}\right)(1 - \beta\mu)
\label{rate}
\end{equation}
where $\Omega=d\mu d\phi$ is the total solid angle
subtended by the source of soft photons, $\mu = \cos{\theta}$, $\sigma$ is
a total cross section for the process, and $dn_{\rm ph}/d\eps/d\Omega$ 
is the local density of the soft photons.

The properties of the field of soft photons are usually simplified by
taking $dn_{\rm ph}/d\eps/d\Omega$ as for the blackbody radiation.
This simplification should be taken with care since magnetised atmospheres
of neutron stars introduce
strong anisotropy as well as spectral distortions to the outgoing radiation \cite{pavlov}.
Effectively it means that the ICS effects obtained with this simplification are just upper limits
to the actual effects. 

To estimate electron cooling rate $\dot{\gamma}_{\sss \rm ICS}$  due to the ICS the differential 
form of (\ref{crosssection}) is necessary:
\begin{eqnarray}
\lefteqn{\frac{d\sigma}{d\Omega^\prime_s} = \frac{3\st}{16\pi}\,
\left[
(1 - {\mu^\prime}^2)(1 - {\mu^\prime_s}^2)\ + \right.}\nonumber\\
&&\hbox{\hskip12mm}\left.+\ \frac{1}{4}\ (1 + {\mu^\prime}^2)
(1 + {\mu^\prime_s}^2)(g_1 + g_2)\right]
\label{diffsection}
\end{eqnarray}
(eg. \cite{dau89}), where $d\Omega^\prime_s = d\phi^\prime_s d\mu^\prime_s$ is 
an increment of solid angle into which outgoing
photons with energy $\epr_s$ in the electron rest frame are directed. 
The mean electron energy loss rate then reads
\begin{eqnarray}
\lefteqn{\dot{\gamma}_{\sss \rm ICS} = - c\int
d\epsilon \int d\Omega\, \left(\frac{dn_{\rm ph}}{d\epsilon d\Omega}\right)
(1 - \beta\mu)\ \times}\nonumber\\
&&\hbox{\hskip20mm}\times 
  \int
d\Omega^\prime_s
\left(  \frac{d\sigma}{d\Omega^\prime_s}  \right)
(\epsilon_s - \epsilon)
\label{lossrate}
\end{eqnarray}
where $\epsilon_s = \epr_s\gamma(1 + \beta\mu^\prime_s)$ is the scattered
photon energy in the lab frame (e.g.\cite{dermer}). 

The spectrum of magnetic ICS calculated  numerically to model the Vela pulsar 
is shown in Fig.\ref{vela} with thin solid line. The blackbody soft photons originating
at the stellar surface (dotted line) are upscattered by secondary $e^\pm$-pairs at the expense of their
``longitudinal" energy $\gamma_\parallel m c^2$, assumed to remain unchanged during
the burst of synchrotron emission.
Without the ICS component due to the $e^\pm$-pairs the RXTE data for the Vela pulsar would be difficult
to reproduce within the model.
It is worth to note, that the magnetic ICS component due to primary electrons (not shown in 
Fig.\ref{vela}) is energetically insignificant comparing
to the CR component because the case of strong accelerating field was used in this particular model.

\subsection{Photon splitting}
Photon splitting into two photons in the presence of magnetic field $B$  is a third-order
QED process with no energy threshold \cite{adler}. 
The attenuation coefficient, after averaging over the polarization states, reads \cite{splitting}
\begin{equation}
T_{\rm split}(\epsilon) 
\approx \frac{\alpha^3}{10\pi^2} \frac{1}{\lambdabar_{\rm c}} \left(\frac{19}{315}\right)^2 
\left(\frac{B \sin{\theta_{\rm kB}}}{B_{\rm crit}}\right)^6 \,  \epsilon^5 \, \,  {\rm cm}^{-1},
\label{split1}
\end{equation}
(provided $B$ does not exceed $B_{\rm crit}$ substantially)
where $\alpha$ is a fine structure constant, $\epsilon$ is the photon energy in units of $m c^2$,
$\theta_{\rm kB}$ is the angle between photon momentum vector and the local magnetic field.  
The process, therefore, strongly depends on magnetic field strength $B$.

Photon splitting has attracted substantial interest in recent years due to the discovery of
neutron stars with supercritical magnetic fields (i.e. magnetars; see Fig.6) \cite{mereghetti}.
It has been analysed in details by \cite{splitting} and incorporated in a Monte Carlo code
tracing the propagation of electromagnetic cascades in the magnetospheres of high-$B$ pulsars. 
The effect was found to explain satisfactorily the unusual cut-off observed in the gamma-ray spectrum of
B1509-58 (see Sect.3). Generally, it becomes competitive to the magnetic pair creation
for dipolar magnetospheres with $B_{\rm s} \simgreat 0.3 B_{\rm crit}$ \cite{splitting}. The degradation of photon energy
in the course of splitting inhibits also any development electromagnetic cascades.
In consequence, high-$B$ RPP should not emit coherent
radio emission. 
Indeed, there exists a high-$B$ region in the $P-\dot P$ diagram (Figs.1,6) void
of radiopulsars. Even though recently discovered (during The Parkes Multibeam Pulsar survey) two high-$B$ radiopulsars 
\cite{camilo} are located above
the limiting line derived by \cite{splitting_deathline}, the general argument for magnetars expected to be
radio-quiet RPP remains valid \cite{bing1}. 

\section{RPP as sources of UHECR generated beyond the light cylinder}
The hypothesis that cosmic ray events above $\sim 5\times 10^{19}$eV, i.e. above the GZK cutoff,
are due to charged particles accelerated by strongly magnetised neutron stars 
is being kept under consideration  \cite{blandford}\cite{olinto}\cite{sigl} 
apparently for two reasons. First, within Down--Top scenarios
which rely on conventional physics 
a list of classes of objects satisfying a necessary condition to generate such cosmic rays
is rather short according to the appealing Hillas diagram \cite{hillas}. Second, no compelling
breakthrough has been achieved in studying other candidates
in this context, though central engines of AGN or
jets extending from Fanaroff-Riley II radio galaxies \cite{biermann} take rather high positions in the list \cite{ong}.

In the case of neutron stars 
the most promising and natural reservoir of required energy is the rotational energy of the stars,
a point reiterated on numerous occasions, e.g.\cite{miller}, and assumed in most models. 
There are three fundamental features discriminating the models:\\
1) the site of acceleration
with respect to the neutron star,\\ 
2) the mechanism allowing to tap the rotational energy by charged particles,\\ 
3) the nature and origin of the charged particles subject to acceleration.

Simple but sounding arguments supported by observed HE-radiation properties of RPP
make the first point rather clear: the process of particle acceleration should take place
beyond the light cylinder.
Potential advantage of this choice over acceleration process within the light cylinder
is twofold.
First, full potential drop across open field lines $\Delta\Phi_{\rm pc}$ (\ref{toy4})
is available, at least in principle, for particles outside the light cylinder, 
whereas the capability of an accelerator inside the magnetosphere is severely constrained 
by copious formation of electron-positron pairs
which short out the electric field ${\cal E}_\|$ easily, i.e. $\Delta\Phi_\| < \Delta\Phi_{\rm pc}$.
Second, unlike inside the magnetosphere, the acceleration of charged particles
is not limited by any radiation losses -- a point especially important in the context of UHECR.
Therefore, a particle of charge $Ze$ 
reaches then maximal possible energy
\begin{equation}
\label{10} 
E_{\rm max} \approx Ze\,\Delta\Phi_{\rm pc} = 6 \times 10^{19}\, Z\, B_{13}\, P_{\rm ms}^{-2}\, {\rm eV},
\end{equation} 
(where $B_{13} \equiv B_{\rm s}/10^{13}\G$) which is substantially  higher
than the energy $E_{\rm p}$ which can be attained in the accelerating field of either the polar gap or outer gap:
$E_{\rm p} \ll Ze\,\Delta\Phi_\| < E_{\rm max}$.

The idea of UHECR events due to an accelerator or a converter of the rotational 
energy into kinetic energy of particles,  located beyond the light cylinder
has been pursued recently in a couple of diametrically different models. 
Historically the first model considered the fate of 
charged particles injected into the plerionic nebula
powered by a shocked relativistic wind from a central pulsar \cite{bell} \cite{lucek}. 
This model was motivated by the theoretical analysis
of magnetohydrodynamics within the Crab Nebula \cite{rees} \cite{kennel} \cite{coroniti}. 
The acceleration was proposed to occur in the electric
field of the shocked wind. A simple structure of the electric field was derived as induced by 
the relativistic radial wind crossing a toroidal magnetic field (which originates inside
the light cylinder, at the stellar surface) with assumed radial and angular dependence after \cite{rees} and \cite{coroniti}. 
Charged particles entering
the nebula are subject to the $\vec {\cal E} \times \vec B$ drift as well as to the $\nabla B$ drift 
(due to strong inhomogeneity in $\vec B$) along various paths, and may gain energy at the expense of $\vec {\cal E}$ 
before exiting. Since the
electric field is potential here, the net change in the particle energy
does not depend on the path but solely on the points of entry and exit. In particular,   
the maximal gain $\Delta E$ of energy is due to potential difference between the pole (fixed by the rotation
axis of the pulsar) and the equator
(cf. (3) in \cite{lucek}), and actually it equals to $E_{\rm max}$ given by (\ref{10}).
In order for the particle to penetrate the nebula from outside it should be already highly
relativistic, with $E_{\rm init}\sim 10^{15}$eV,  presumably pre-accelerated by diffusive acceleration at the outer shock
where the supernova remnant meets the interstellar medium. Otherwise the Larmor radius of the particle is to too small for
the $\nabla B$ inward drift at the pole (the case relevant for one combination of signs of  $\vec B$ and $Ze$)
to counter the $\vec {\cal E} \times \vec B$ outward drift.
Though very attractive, the model is unable to make any strong predictions about
spectral properties of its UHECR without reliable assumptions about
spatial distribution and directional properties of the pre-accelerated particles.

All other models make use of charged particles, either protons or iron nuclei, coming from within the light cylinder
i.e. supplied by a neutron star itself. \\
A recently proposed phenomenological model of UHECR  within the Galaxy
incorporates 
iron Fe$^{26}$ nuclei ($Z = 26\, Z_{26}$) accelerated in a relativistic MHD wind flowing out of 
very young, rapidly rotating highly magnetised neutron stars \cite{blasi}.
The number of nuclei crossing the light cylinder per unit time is taken equal to the Goldreich--Julian rate 
estimated at the polar cap
\begin{equation}
\label{11} 
\dot N({\rm Fe}^{26}) \approx \dot N_{\rm GJ} = A_{\rm pc} \, \frac{\varrho_{\rm GJ}}{Z e \, c}, 
\end{equation}
where $A_{\rm pc} \simeq \pi R^3_{\rm s}  R_{\rm lc}^{-1}$ and $\varrho_{\rm GJ}$ is defined in (\ref{4}).
It is also assumed that electron-positron pairs formed within the magnetosphere do not dominate the flow in terms of 
the rest mass. This is a reasonable assumption because
otherwise the number of created e$^\pm$ pairs per nucleus would have to exceed $m_{\rm Fe}/ 2 m_{\rm e} \approx 5\times 10^4$ 
which is unlikely in very strong magnetic fields $B_{\rm s} > 10^{13}$G due to photon-splitting
effects \cite{splitting}.
A key postulate of the model is that transfer of a sizeble fraction ($\xi \simless 1$) of the spindown luminosity occurs 
just beyond the light cylinder into the kinetic energy flux of the ions. 
At the light cylinder, however,
the spindown luminosity (i.e. the rotational energy loss rate) is usually thought to be dominated by
the Poynting flux (of low-frequency electromagnetic waves).  
In terms of the so called magnetization parameter $\sigma$  which
is defined as the ratio of the Poynting flux to the the particle kinetic energy flux \cite{rees} \cite{coroniti}
it means that -- whatever the reason -- $\sigma$
does not remain constant in the wind:
$\sigma \gg 1$ at the light cylinder converts further out to $\sigma \ll 1$, i.e. the wind must depart from
an ideal MHD flow case.
\begin{figure}
\begin{center}
\includegraphics[width=1.3\textwidth]{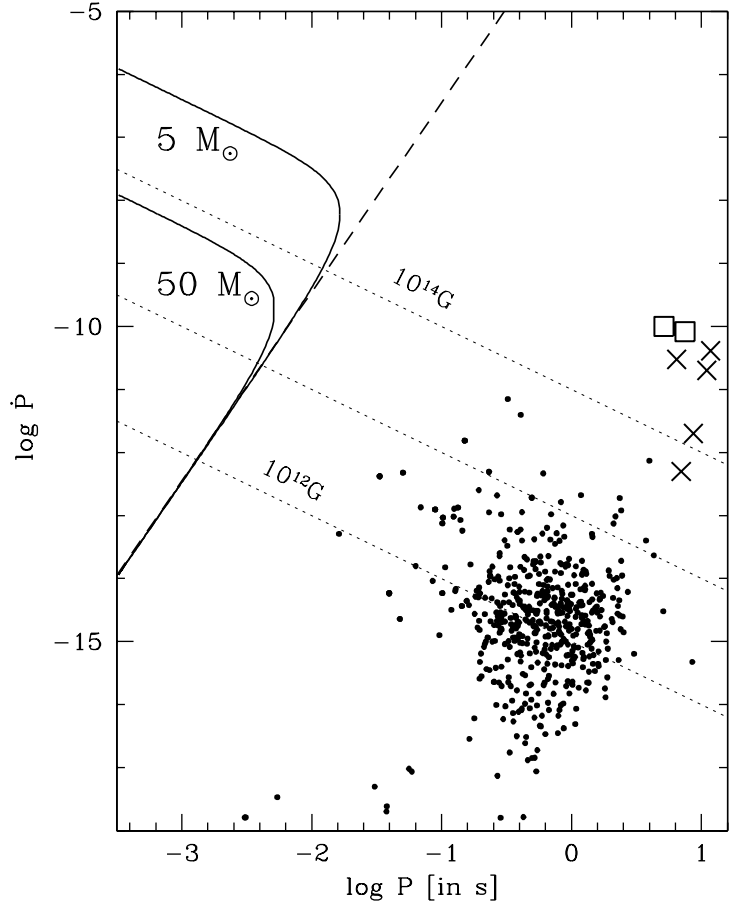}
\end{center}
\vspace*{-1.7cm}
\caption[]{The high--$B_{\rm s}$ part of the $P - \dot P$ diagram (see Fig.\ref{ppdot}).
Dashed line indicates the full potential drop $\Delta\Phi_{\rm pc}$ across open field lines
equal to $\frac{1}{26}\times 10^{20}$V.
The allowed region in the model of \cite{olinto} 
to accelerate a fully ionized atom of iron ($Z = 26$) to the energy of $10^{20}$eV and let it
traverse the pre-supernova envelope without spallation effects
lies to the left of two solid lines which are drawn for two values of the mass of the envelope: $5 M_\odot$ and $50 M_\odot$.
A group of magnetars, clustered around $P \sim \, 10$s is also indicated after \cite{mereghetti}
although their X-ray activity is not driven
by rotational energy losses (i.e. they do not belong to the class of RPP) but rather involves accretion
or the decay of strong magnetic field. 
Open squares and crosses denote SGR and AXP, respectively.}
\label{eps2}
\end{figure}  
This picture
is motivated again by the wind models for the Crab Nebula (e.g. \cite{emmering} \cite{kennel}, see also \cite{shibata}
for calorimetric properties of the Crab Nebula, and \cite{arons98} for an account of the problem
of the coupling of RPP to plerionic nebulae). No consensus about likely mechanisms responsible for the dissipation
of the Poynting flux has been reached so far,
though several models have been proposed (e.g. \cite{coroniti} \cite{begelman}).
For a single average iron nucleus the postulated conversion means
acceleration up to 
\begin{equation}
\label{12}
E_{\rm UHECR} \approx L_{\rm sd} / \dot N({\rm Fe}^{26}) \approx 
Z e \,\Delta \Phi_{\rm pc} \approx 10^{21} Z_{26}\, B_{13}\, P^{-2}_{\rm ms}\, {\rm eV},
\end{equation}
where (\ref{11}) was used.
Similarly as in the previous model, the energy gain 
is in fact directly related to the full potential drop $\Delta\Phi_{\rm pc}$ across open field lines.
For a fully ionized iron ($Z_{26}= 1$) to reach $E_{\rm UHECR} = 10^{20}$eV
the pulsar would have to be fastly spinning and highly magnetised (for example
$P = 0.001$s and $B_{\rm s} = 10^{12}$G). According to \cite{blasi}, 
the region in the $P-B_{\rm s}$ space occupied by pulsars 
satisfying the requirement $E_{\rm UHECR} = 10^{20}$eV is actually even more constrained then (\ref{12})
suggests. The magnetic field of the pulsar should not be too strong to make sure that the characteristic
time scale of spin-down $\tau$
is not too short allowing thus the expanding pre-supernova envelope of mass $M_{\rm env}$ to disperse and
become transparent so the iron nuclei would escape it without significant spallation. 
For $M_{\rm env}$ equal $5 M_\odot$ and $50 M_\odot$
the magnetic field $B_{\rm s}$ should not exceed $\sim 6\times 10^{13}$G and $\sim 6\times 10^{14}$G, respectively
(see Fig.\ref{eps2}).
A factor not discussed in \cite{blasi} but likely to further constrain the allowed region
in the $B_{\rm s}-P$ space in order to 
make the model work is a poorly known cohesive or binding energy $\Delta \varepsilon_c$ of iron 
on the high-$B_{\rm s}$ neutron star surface.
For $B_{\rm s} > 10^{13}\G$ $\Delta \varepsilon_c$ may exceed $5\keV$ \cite{abrahams}.
Thermionic emission of iron nuclei is then impossible unless
the surface temperature exceeds $\sim 3\times 10^6$K within the first $\sim 10^8$s \cite{nomoto} after the neutron star formation
which is a characteristic time scale mediated by the neutron star's spin-down when the energy $E_{\rm max}$ 
drops below a required CR energy $\sim 10^{20}$eV, and the expansion of the envelope during which it becomes transparent to
the Fe nuclei.
The only supply of iron nuclei may then occur via field emission. 
Making modest assumptions about the distribution of rapidly rotating neutron stars, the rates of their birth
and -- most importantly -- the efficiency rate for acceleration of the particles at the light cylinder, 
an expected flux of these particles on Earth is compared to the data. 
The calculated particle spectrum  above $5\times 10^{19}$eV
is flat, $N_{\rm Fe} \propto E^{-\gamma}$ with $\gamma = 1$, and therefore should be visible as a distinct 
component in the CR spectrum, which is much steeper below $5\times 10^{19}$eV with the power-law index $\gamma \approx 3$.  
However, the number of events above $5\times 10^{19}$eV detected so far is insufficient for describing
them in terms of spectral features \cite{cronin}.

For a pulsar capable of producing UHECR as in \cite{blasi}, a natural continuation --
as it slows down its rotation  and becomes a Crab-like object -- would be to generate CR particles of lower energy. 
Hadronic model proposed by \cite{bednarek} seems to be relevant in this context, since it describes the fate
of iron nuclei extracted from the pulsar surface and then subject to acceleration.
However, a different acceleration site is considered here: the nuclei are assumed to accelerate within the light cylinder,  
in outer gaps of \cite{ho} (i.e. the type of accelerator not addressed in this review). Once the nuclei achieve
high energy ($\gamma_{\rm Fe} > 10^5$) they disintegrate in the field of soft photons present in the outer gap. 
Relativistic neutrons extracted in this way decay into protons either inside or outside the surrounding nebula. 
In the first case, these relativistic protons interact with nebular material via $pp$ processes giving rise
to VHE photons and neutrinos. In the second case, cosmic ray
protons peaking at $\sim 10^{15} - 10^{16}\eV$ (close to the knee in the CR spectrum) are produced. 

A distinct model of UHECR particles above the GZK cutoff has been recently proposed \cite{degouveia}.
It postulates that UHECR are protons accelerated in reconnection sites above the magnetospheres of 
very young pulsars formed by accretion-induced collapse of white dwarves (AIC).  
Promising candidates would occupy roughly the same region of the $P-\dot P$ space 
as indicated for the model of \cite{blasi} (the less massive
envelope case) in Fig.\ref{eps2}. Since the estimated rate of AIC events per galaxy is too low to rely just on our Galaxy,
the contribution of all galaxies within the distance of about 50~Mpc is necessary to reach the observed rate
of UHECR events.

\begin{figure}
\begin{center}
\includegraphics[width=0.8\textwidth]{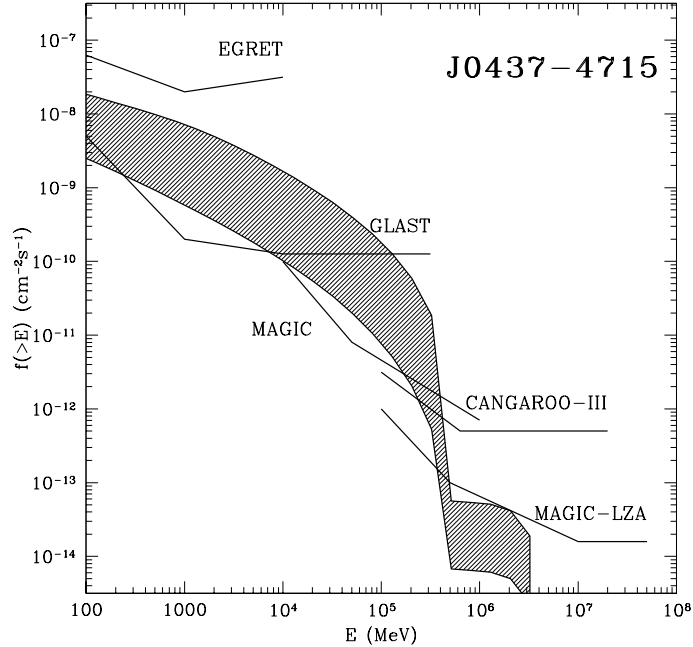}
\end{center}
\caption[]{Cummulative~spectral flux of photons expected for the millisecond pulsar
J0437-4715 at a distance of $140\,$pc according to \cite{bulik}. 
The shaded region shows the range of flux levels due to uncertainity
in the maximal energy of primary electrons. The main part of the spectrum is due
to curvature radiation of the electrons. The additional feature reaching the VHE
domain is due to inverse Compton scattering of soft photons 
from the surface with the temperature $4\times 10^5$K.
Sensitivities of EGRET as well as three major HE and VHE 
experiments of the future are also indicated. MAGIC--LZA denotes
sensitivity of MAGIC in its Large Zenith Angle mode.
}
\label{j0437}
\end{figure}

\section{Concluding remarks}
High energy astrophysics of neutron stars 
received an impressive boost from two major satellite missions of the past -- ROSAT and CGRO, backed by 
still ongoing experiments -- ASCA, BeppoSAX and RXTE.
Theoretical astrophysics of RPP was confronted with--  and surprised by
an unprecedented variety of spectral and temporal properties among the detected sources.
Another unexpected challenge came from radio-astronomy, due to superb performance of The Parkes Multibeam Pulsar Survey,
with recent discoveries of radiopulsars,
with extremely high magnetic fields in two cases and an extremely long spin period in another case.
Numerous modifications (both, minor and major) to the existing models of magnetospheric activity
were invented to accommodate at least some of these properties.
Several predictions have been presented which would hopefully discriminate between those models.
It will be impossible, however, to verify those predictions without achieving higher sensitivities 
and exploring new energy domains.

A break-through in understanding
rotation powered pulsars and their plerionic environments should then come from HE/VHE astronomy of the near future,
with its planned satellite and ground-based experiments. Expected sensitivity and energy range
for some of them is presented in Fig.\ref{j0437} along with a predicted flux from a nearby millisecond pulsar,
overlaid for the sake of comparison.
The satellite experiment GLAST \cite{glast}
will be superior to its predecessor -- EGRET on board CGRO -- in two
aspects. First, its sensitivity at 10~GeV will be three orders of magnitude better than
that of EGRET. Second, it will reach energy of 300~GeV, closing thus for the first time
a wide gap in energy between ground-based and satelite experiments.
The MAGIC Telescope \cite{blanch} -- a 17 m diameter Imaging Air Cherenkov Telescope  (IACT) -- is expected to 
operate with sensitivity about three orders of magnitude higher at 10~GeV
than EGRET. Its advanced technology will make 
possible to cover energy range  between 10~GeV and 1~TeV, and to reach $\sim 50\,$TeV in the
Large Zenith Angle mode. 
Energy ranges of GLAST and MAGIC will overlap over more than one decade in energy.
Another proposed IACT, VERITAS \cite{veritas}, will be an array of seven 10~m telescopes,
covering energy range from 50~GeV to 50~TeV with planned sensitivity at 1~TeV about ten times
better than MAGIC.

\section*{Acknowledgments}

I am grateful to J.~Arons, W.~Bednarek, T.~Bulik, K.S.~Cheng, J.~Dyks, E.M.~de~Gouveia~Dal~Pino, 
M.~Gros, P.~Haensel, O.C.~de~Jager,
A.V.~Olinto, S.~Shibata, T.C.~Weekes,
and B.~Zhang for sharing with me their views and opinions on 
issues addressed in this review. Financial support by KBN grant 2P03D02117 is acknowledged.

\end{document}